# JOSEPHSON CURRENTS IN POINT CONTACTS BETWEEN DIRTY TWO-BAND SUPERCONDUCTORS

Y.S. Yerin, A.N. Omelyanchouk

*B.Verkin Institute for Low Temperature Physics and Engineering
of the National Academy of Sciences of Ukraine, 47 Lenin Ave., Kharkov 61103 , Ukraine*

yerin@ilt.kharkov.ua

**Abstract**

We developed microscopic theory of Josephson effect in point contacts between dirty two-band superconductors. The general expression for the Josephson current, which is valid for arbitrary temperatures, is obtained. This expression was used for calculation of current-phase relations and temperature dependences of critical current with application to $MgB_2$ superconductor. Also we have considered influence on contact characteristics interband scattering effect appeared in case of dirty superconductors. It is shown that the correction to Josephson current due to the interband scattering depends on phase shift in the banks (i.e. s- or s±-wave symmetry of order parameters).

## 1. INTRODUCTION

The first mentioning about multiband superconductivity has appeared in theoretical works of Matthis, Suhl and Walker [1] and Moskalenko and Palistrant [2] more than 50 years ago. At that time their papers were considered as attempts to fit for BSC-theory some characteristics of superconducting materials (refinement of the ratio of an energy gap to critical temperature, heat capacity jump, london penetration depth etc.). Really multiband superconductivity became hot topic in condensed matter physics in 2001, when Nagamatsu has discovered two-band superconductivity in $MgB_2$ with anomalous high $T_c=39$ K [3]. It's striking that pairing mechanism had electron-phonon origin in magnesium diboride and order parameters, which attribute to superconducting energy gaps, have s-wave symmetry.

Iron-based superconductors, which have discovered not long ago [4], most probably can be adding on to multiband systems. For example, ARPES specifies on existence of two full gaps in $Ba_{0.6}K_{0.4}Fe_2As_2$ [5]. Moreover there is an assumption that in this iron-based superconductor π-shift between phases of order parameters and thus so called extended with sign-reversal of order parameter or s±-wave symmetry is realized [6]. Existence of such phase-shift can lead to new fundamental phenomena and effects in these superconducting systems. Unfortunately the most of modern experimental methods (measuring magnetic penetration length, calorimetric method, Knight shift and spin-relaxation velocity in NMR and above mentioned ARPES) cannot give unique answer about gap pairing symmetry. However, recently new phase-sensitive technique based on proximity effect between niobium wire and massive $NdFeAsO_{0.88}F_{0.12}$ plate for probing unconventional pairing symmetry in the iron pnictides was reported [7].

Therefore it's interesting to research phase coherent effects in two-band superconductors which are sensible to shift of phase. Josephson effects are one of such sensitive to the phase phenomena. At the present time there are many theoretical publications, devoted to this problem [8-12]. In these papers the Josephson effect in tunnel junctions between one-band and two-band superconductors is considered.

In present paper we study the stationary Josephson effect in S-C-S (superconductor-constriction-superconductor) contact, which behavior even in the case of one-band superconductors, as was revealed in Kulik and Omelyanchouk papers [13,14] (KO theory), has the qualitative differences with respect to S-I-S (superconductor-insulator-superconductor) tunnel junctions. We built the microscopic theory of the «dirty» S-C-S contact for two-band superconductors, which generalizes the KO theory for this case. We show that the interband scattering effects occurring in dirty superconductors result in mixing of Josephson currents between different bands.

## 2. MODEL AND BASIC EQUATIONS

Consider the weak superconducting link as a thing filament of length *L* and diameter *d*, connecting two bulk superconductors (banks) (Fig.1). Such model describes the S-C-S (Superconductor-Constriction-Superconductor) contacts with direct conductivity (point contacts, microbridges), which qualitatively differ from the tunnel S-I-S junctions. On condition that $d \ll L$ and $d \ll \min[\xi_1(0), \xi_2(0)]$ ($\xi_i(T)$ - coherence lengths) we can solve inside the filament ($0 \leq x \leq L$) a one-dimensional problem with "rigid" boundary conditions. At $x = 0, L$ all functions are assumed equal to the values in homogeneous no-current state of corresponding bank.

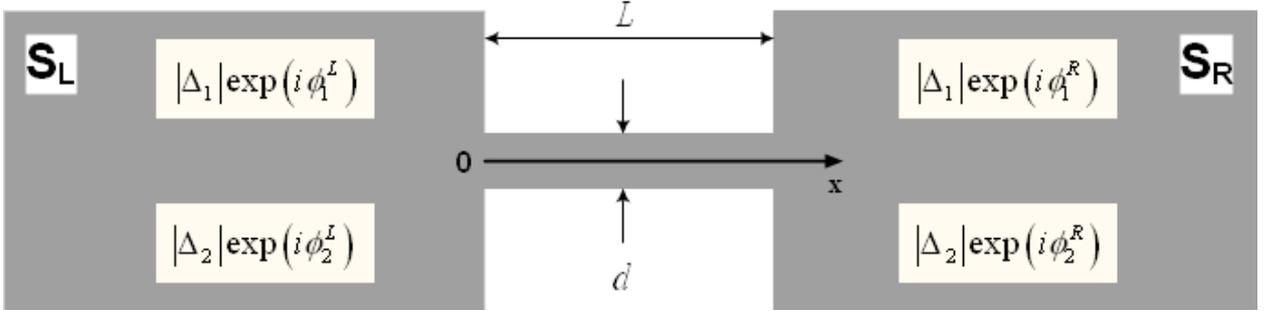

Fig.1. Model of Superconductor-Constriction-Superconductor contact. The right and left banks are bulk two-band superconductors connected by the thing filament of length L and diameter d.

We investigate the case of dirty two-band superconductor with strong impurity intraband scattering rates (dirty limit) and weak interband scattering. In the dirty limit superconductor is described by the Usadel equations for normal and anomalous Green's functions $g$ and $f$, which for two-band superconductor take the form [15]:

$$\omega f_1 - D_1\left(g_1 \nabla^2 f_1 - f_1 \nabla^2 g_1\right) = \Delta_1 g_1 + \gamma_{12}\left(g_1 f_2 - g_2 f_1\right), \qquad (1)$$

$$\omega f_2 - D_2\left(g_2 \nabla^2 f_2 - f_2 \nabla^2 g_2\right) = \Delta_2 g_2 + \gamma_{21}\left(g_2 f_1 - g_1 f_2\right). \qquad (2)$$

Usadel equations are supplemented with self-consistency equations for order parameters $\Delta_i$:

$$\Delta_i = 2\pi T \sum_j \sum_{\omega > 0}^{\omega_D} \lambda_{ij} f_j, \qquad (3)$$

and with expression for the current density

$$j = -ie\pi T \sum_i \sum_\omega N_i D_i \left(f_i^* \nabla f_i - f_i \nabla f_i^*\right). \qquad (4)$$

Index $i = 1,2$ numerates the first and the second bands. Normal and anomalous Green's functions $g_i$ and $f_i$, which are connected by normalization condition $g_i^2 + |f_i|^2 = 1$, are functions of $x$ and the Matsubara frequency $\omega = (2n+1)\pi T$. $D_i$ are the intraband diffusivities due to nonmagnetic

impurity scattering, $N_i$ are the density of states on the Fermi surface of the $i$-th band, electron-phonon constants $\lambda_{ij}$ take into account Coulomb pseudopotentials and $\gamma_{ij}$ are the interband scattering rates. There are the symmetry relations $\lambda_{12} N_1 = \lambda_{21} N_2$ and $\gamma_{12} N_1 = \gamma_{21} N_2$.

In considered case of short weak link we can neglect all terms in the equations (1), (2) except the gradient one. Using the normalization condition we have equations for $f_{1,2}$

$$\sqrt{1-|f_1|^2}\, \frac{d^2}{dx^2} f_1 - f_1 \frac{d^2}{dx^2} \sqrt{1-|f_1|^2} = 0, \tag{5}$$

$$\sqrt{1-|f_2|^2}\, \frac{d^2}{dx^2} f_2 - f_2 \frac{d^2}{dx^2} \sqrt{1-|f_2|^2} = 0. \tag{6}$$

The boundary conditions for equations (5) and (6) are determined by solutions of equations for Green's functions in left (right) banks:

$$\begin{cases} \omega f_1^{L(R)} = \Delta_1^{L(R)} \sqrt{1-\left|f_1^{L(R)}\right|^2} + \gamma_{12}\left(\sqrt{1-\left|f_1^{L(R)}\right|^2}\, f_2^{L(R)} - \sqrt{1-\left|f_2^{L(R)}\right|^2}\, f_1^{L(R)}\right), \\ \omega f_2^{L(R)} = \Delta_2^{L(R)} \sqrt{1-\left|f_2^{L(R)}\right|^2} + \gamma_{21}\left(\sqrt{1-\left|f_2^{L(R)}\right|^2}\, f_1^{L(R)} - \sqrt{1-\left|f_1^{L(R)}\right|^2}\, f_2^{L(R)}\right). \end{cases} \tag{7}$$

Note, that solutions $f_1$ and $f_2$ of equations (5, 6) are coupled due to the interband scattering in the banks through the boundary conditions (7).

Introducing the phases of order parameters in banks

$$\Delta_1^{L(R)} = |\Delta_1| \exp(i\phi_1^{L(R)}),\ \Delta_2^{L(R)} = |\Delta_2| \exp(i\phi_2^{L(R)}), \tag{8}$$

and writing $f_i(x)$ in equations (5) and (6) as $f_i(x) = |f_i(x)| \exp(i\chi_i(x))$ we have

$$|f_i(0)| = |f_i(L)| = |f_i|,\ \chi_i(0) = \chi_i^L,\ \chi_i(L) = \chi_i^R, \tag{9}$$

where $|f_i|$ and $\chi_i^{L(R)}$ are connected with $|\Delta_i|$ and $\phi_i^{L(R)}$ through equations (7).

The solution of equations (5-9) determines the Josephson current in the system. It depends on the phase difference on the contact, which we define as $\phi \equiv \phi_1^R - \phi_1^L = \phi_2^R - \phi_2^L$, and from possible phase shift in each banks $\delta = \phi_1^L - \phi_2^L = \phi_1^R - \phi_2^R$. The phase shift $\delta$ between the phases of the two order parameters in two-band superconductor can be 0 or $\pi$, depending on the sign on the interband coupling constants, the values of the interband scattering rates and the temperature of the system (see Appendix).

Equations (5), (6) with boundary conditions (9) admit analytical solution, and for the current density (4) we obtain:

$$I = \frac{2\pi T}{eR_{N1}} \sum_{\omega} \frac{|f_1|\cos\frac{\chi_1^L - \chi_1^R}{2}}{\sqrt{(1-|f_1|^2)\sin^2\frac{\chi_1^L - \chi_1^R}{2} + \cos^2\frac{\chi_1^L - \chi_1^R}{2}}}$$

$$\times \arctan \frac{|f_1|\sin\frac{\chi_1^L - \chi_1^R}{2}}{\sqrt{(1-|f_1|^2)\sin^2\frac{\chi_1^L - \chi_1^R}{2} + \cos^2\frac{\chi_1^L - \chi_1^R}{2}}}$$

$$+ \frac{2\pi T}{eR_{N2}} \sum_{\omega} \frac{|f_2|\cos\frac{\chi_2^L - \chi_2^R}{2}}{\sqrt{(1-|f_2|^2)\sin^2\frac{\chi_2^L - \chi_2^R}{2} + \cos^2\frac{\chi_2^L - \chi_2^R}{2}}}$$

$$\times \arctan \frac{|f_2|\sin\frac{\chi_2^L - \chi_2^R}{2}}{\sqrt{(1-|f_2|^2)\sin^2\frac{\chi_2^L - \chi_2^R}{2} + \cos^2\frac{\chi_2^L - \chi_2^R}{2}}}. \quad (10)$$

Here $R_{N1}$, $R_{N2}$ - contributions to normal resistance of each band, where $R_{Ni}^{-1} = \frac{2Se^2}{L}N_iD_i$ ($S = \frac{\pi d^2}{4}$ is cross section area).

The general expression (10), together with Eqs. (7-9) describes the Josephson current as function of gaps in the banks $|\Delta_i|$ and phase difference on the contact $\phi$.

### 3. JOSEPHSON CURRENTS

#### 3.1 Josephson current without interband scattering

When interband scattering is vanished, from system of equations (7) we obtain:

$$\begin{cases} f_{1,2}^{L(R)} = \dfrac{\Delta_{1,2}^{L(R)}}{\sqrt{\left|\Delta_{1,2}^{L(R)}\right|^2 + \omega^2}} \\ g_{1,2}^{L(R)} = \dfrac{\omega}{\sqrt{\left|\Delta_{1,2}^{L(R)}\right|^2 + \omega^2}} \end{cases} \quad (11)$$

Taking into account (11) we rewrite expression for the current (10) in terms of order parameters:

$$I = \frac{2\pi T}{eR_{N1}} \sum_\omega \frac{|\Delta_1|\cos\frac{\phi}{2}}{\sqrt{\omega^2 + |\Delta_1|^2 \cos^2\frac{\phi}{2}}} \arctan \frac{|\Delta_1|\sin\frac{\phi}{2}}{\sqrt{\omega^2 + |\Delta_1|^2 \cos^2\frac{\phi}{2}}}$$
$$+ \frac{2\pi T}{eR_{N2}} \sum_\omega \frac{|\Delta_2|\cos\frac{\phi}{2}}{\sqrt{\omega^2 + |\Delta_2|^2 \cos^2\frac{\phi}{2}}} \arctan \frac{|\Delta_2|\sin\frac{\phi}{2}}{\sqrt{\omega^2 + |\Delta_2|^2 \cos^2\frac{\phi}{2}}}, \quad (12)$$

Thus, if to neglect interband scattering rates $\gamma_{ik}$ Josephson current (10) decomposes on two parts: current flows independently from the first band to the first one and from the second band to the second one. For zero temperature $T = 0$ expression (12) takes the form

$$I = \frac{\pi|\Delta_1|}{2eR_{N1}} \cos\frac{\phi}{2} \operatorname{Ar\,tanh} \sin\frac{\phi}{2} + \frac{\pi|\Delta_2|}{2eR_{N2}} \cos\frac{\phi}{2} \operatorname{Ar\,tanh} \sin\frac{\phi}{2} \quad (13)$$

The expression (12) is straightforward generalization of Josephson current for one-band superconductor [13]. The partial inputs of bands currents $I(1 \to 1)$ and $I(2 \to 2)$ are determined by the ratio of normal resistances $r = \dfrac{R_{N1}}{R_{N2}}$. Introducing the total resistance $R_N = \dfrac{R_{N1}R_{N2}}{R_{N1} + R_{N2}}$ and normalizing the current on the value $I_0 = \dfrac{2\pi}{eR_N}T_c$ we plot on Fig.2 the current-phase relation (12) for different values of $r$ and temperature $T$.

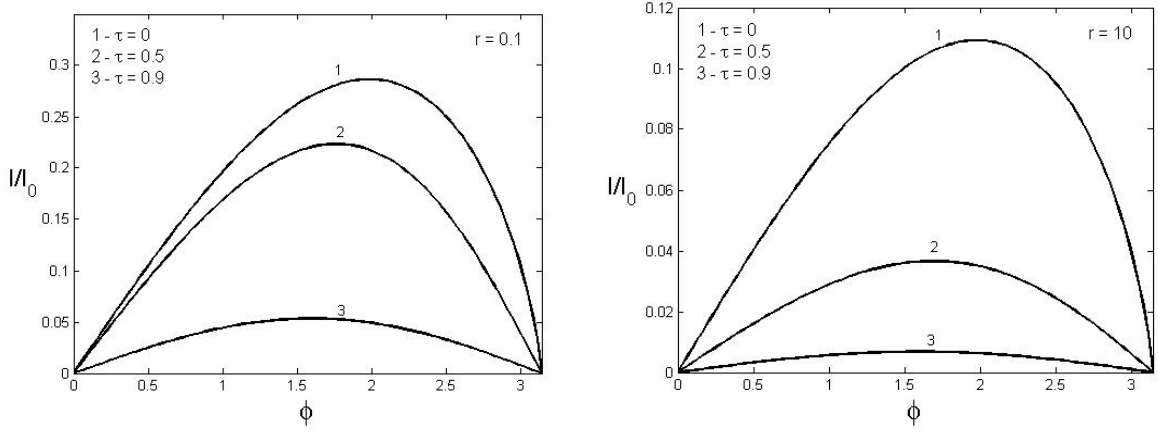

Fig.2. Current-phase relations of S-C-S MgB$_2$|MgB$_2$ for different temperatures $\tau = T/T_c$ and ratio of resistances $r = \dfrac{R_{N1}}{R_{N2}}$.

The current-phase relation (12) determines the temperature dependence of critical current $I_c$, which is plotted on Fig.3.

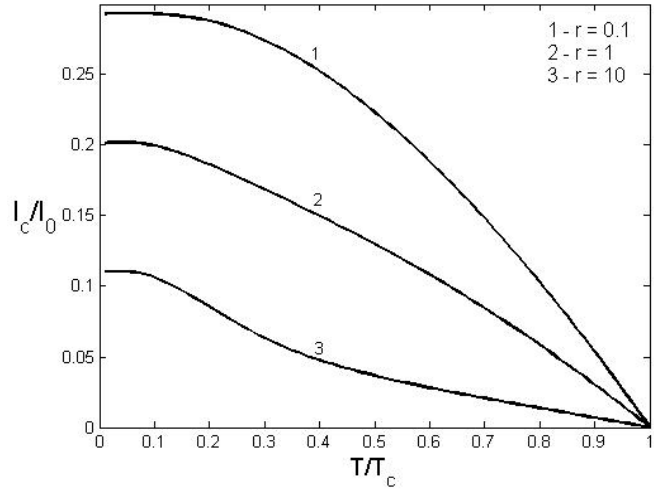

Fig.3. Temperature dependences of critical current $I_c(T)$ for different values of $r = \dfrac{R_{N1}}{R_{N2}}$.

In calculations of $I(\phi)$ and $I_c(T)$ (Fig.2,3) we use the parameters for superconductor MgB$_2$ with $\gamma_{ij} = 0$. In the case of zero interband scattering the Josephson current (12) does not depend on possible phase shift $\delta$ (i.e. in the case of s±-wave symmetry).

### 3.2 Josephson current with interband scattering

Now our aim is investigation of effects of interband scattering on S-C-S contact behavior. In general, the system (7) has no analytical solution. The case of temperatures near critical temperature $T_c$ and arbitrary strength of interband scattering was considered in [16]. Here, in the case of arbitrary temperature $0 \leq T \leq T_c$ we consider the interband scattering $\gamma_{ij}$ by the theory of perturbations. In the first approximation for Green's functions in each bank we obtain:

$$\begin{cases} f_1 = \dfrac{\Delta_1}{\sqrt{|\Delta_1|^2 + \omega^2}} + \gamma_{12} \dfrac{\left(2\omega^2 + |\Delta_1|^2\right)(\Delta_2 - \Delta_1) - \Delta_1^2\left(\Delta_2^* - \Delta_1^*\right)}{2\sqrt{\left(|\Delta_1|^2 + \omega^2\right)^3}\sqrt{|\Delta_2|^2 + \omega^2}}, \\ f_2 = \dfrac{\Delta_2}{\sqrt{|\Delta_2|^2 + \omega^2}} + \gamma_{21} \dfrac{\left(2\omega^2 + |\Delta_1|^2\right)(\Delta_1 - \Delta_2) - \Delta_2^2\left(\Delta_1^* - \Delta_2^*\right)}{2\sqrt{|\Delta_1|^2 + \omega^2}\sqrt{\left(|\Delta_2|^2 + \omega^2\right)^3}}. \end{cases} \quad (14)$$

When the interband scattering is taken into account and if phase shift $\delta \neq 0$, the phases of Green functions $f_i$ not coincide with phases of order parameters $\Delta_i$.

Using (14) we obtain expressions for the corrections to the current (12):

$$\delta I = \delta I_1 + \delta I_2, \quad (15)$$

$$\delta I_1 = \frac{2\pi T \gamma_{12}}{eR_{N1}} \sum_\omega \left( \frac{\omega^2\left(|\Delta_2|e^{i\delta} - |\Delta_1|\right)\cos\left(\frac{\phi}{2}\right)}{\sqrt{\left(|\Delta_1|^2 \cos^2\left(\frac{\phi}{2}\right) + \omega^2\right)^3}\sqrt{|\Delta_2|^2 + \omega^2}} \arctan \frac{|\Delta_1|\sin\frac{\phi}{2}}{\sqrt{\omega^2 + |\Delta_1|^2 \cos^2\frac{\phi}{2}}} \right.$$

$$\left. + \frac{1}{2} \frac{\omega^2 |\Delta_1|\left(|\Delta_2|e^{i\delta} - |\Delta_1|\right)\sin\phi}{\left(|\Delta_1|^2 + \omega^2\right)\left(|\Delta_1|^2 \cos^2\left(\frac{\phi}{2}\right) + \omega^2\right)\sqrt{|\Delta_2|^2 + \omega^2}} \right) \quad (16)$$

$$\delta I_2 = \frac{2\pi T \gamma_{21}}{eR_{N2}} \sum_\omega \left( \frac{\omega^2\left(|\Delta_1| - e^{i\delta}|\Delta_2|\right)\cos\left(\frac{\phi}{2}\right)}{\sqrt{\left(|\Delta_2|^2 \cos^2\left(\frac{\phi}{2}\right) + \omega^2\right)^3}\sqrt{|\Delta_1|^2 + \omega^2}} \arctan \frac{|\Delta_2|\sin\frac{\phi}{2}}{\sqrt{\omega^2 + |\Delta_2|^2 \cos^2\frac{\phi}{2}}} \right.$$

$$\left. + \frac{1}{2} \frac{\omega^2 |\Delta_2|\left(|\Delta_1| - e^{i\delta}|\Delta_2|\right)\sin\phi}{\left(|\Delta_2|^2 + \omega^2\right)\left(|\Delta_2|^2 \cos^2\left(\frac{\phi}{2}\right) + \omega^2\right)\sqrt{|\Delta_1|^2 + \omega^2}} \right) \quad (17)$$

The correction to Josephson current due to the interband scattering (15-17) depends on phase shift in the banks $\delta = 0$ or $\pi$. This reflects the mixing of Josephson currents between different bands (see [16]).

**CONCLUSIONS**

We developed microscopic theory of Josephson effect in point contacts between dirty two-band superconductors. The general expression for the Josephson current, which is valid for arbitrary temperatures, is obtained. We considered the dirty superconductors with interband scattering effect. This effect in the contacting superconductors produces the coupling of the currents between different bands. With taking into account the interband scattering the observable characteristics $I(\phi)$ and $I_c(T)$ depend on the phase shift of the order parameters in different bands. It permits to distinguish between the $s$ and $s\pm$ wave symmetry by studying the Josephson effect in point contacts of two-band superconductors.

**Appendix**

Free energy within microscopic consideration is given by [17]:

$$F = \frac{1}{2}\sum_{ij}\Delta_i\Delta_j^* N_i \lambda_{ij}^{-1} + F_1 + F_2 + F_{int} + \frac{B^2}{8\pi}, \text{ where} \qquad (A1)$$

$$F_i = 2\pi T \sum_{\omega>0} N_i \left[ (1-g_i) - \text{Re}(f_i^*\Delta_i) + \frac{1}{4}D_i\left(\left(\nabla + \frac{2\pi i \vec{A}}{\Phi_0}\right)f_i\left(\nabla - \frac{2\pi i \vec{A}}{\Phi_0}\right)f_i^* + (\nabla g_i)^2\right)\right], \quad (A2)$$

$$F_{int} = 2\pi T \, \text{Re} \sum_{\omega>0} (N_1\gamma_{12} + N_2\gamma_{21})(g_1 g_2 + f_1^* f_2 - 1). \qquad (A3)$$

Representing order parameters as $\Delta_i^{L(R)} = |\Delta_i|\exp(i\phi_i^{L(R)})$ and anomalous Green functions $f_i(x) = |f_i(x)|\exp(i\chi_i(x))$ and simultaneously taking into account that $N_1\lambda_{12} = N_2\lambda_{21}$ and $N_1\gamma_{12} = N_2\gamma_{21}$ we extract from (A1) terms, which contain the phase difference $\delta = \phi_2^{L(R)} - \phi_1^{L(R)}$:

$$-\frac{\lambda_{12}|\Delta_1||\Delta_2|}{\lambda_{11}\lambda_{22} - \lambda_{12}\lambda_{21}}\cos\delta + 4\pi T\gamma_{12}\sum_{\omega>0}|f_1||f_2|. \qquad (A4)$$

Substitution of $|f_1|$ and $|f_2|$ into (A4) assuming $\gamma_{12} \to 0$ yields

$$\left(-\frac{\lambda_{12}}{\lambda_{11}\lambda_{22} - \lambda_{12}\lambda_{21}} + 4\pi T\gamma_{12}\sum_{\omega>0}\frac{1}{\sqrt{(\omega^2+|\Delta_1|^2)(\omega^2+|\Delta_2|^2)}}\right)\cos\delta. \qquad (A5)$$

For $\delta = 0$ minimum of free energy is satisfied if bracketed expression is less than zero and vice versa for $\delta = \pi$ the minimum of (A1) takes place if bracketed expression is greater than 0. On the basis of this we obtain existence criterion of π-shift between phases of order parameters:

$$\text{sgn}\left(-\frac{\lambda_{12}}{\lambda_{11}\lambda_{22}-\lambda_{12}\lambda_{21}}+4\pi T\gamma_{12}\sum_{\omega>0}\frac{1}{\sqrt{\left(\omega^2+|\Delta_1|^2\right)\left(\omega^2+|\Delta_2|^2\right)}}\right)=\pm 1. \quad (A6)$$

In this condition "+1" correspond to $\delta=\pi$ and "-1" takes place if $\delta=0$.
The derived criterion can be simplified at $T=0$:

$$\text{sgn}\left(-\frac{\lambda_{12}}{\lambda_{11}\lambda_{22}-\lambda_{12}\lambda_{21}}+\frac{2\gamma_{12}}{|\Delta_2(0)|}\text{K}\left(\sqrt{1-\frac{|\Delta_1|^2(0)}{|\Delta_2|^2(0)}}\right)\right)=\pm 1, \quad (A7)$$

where $\text{K}\left(\sqrt{1-\frac{|\Delta_1|^2(0)}{|\Delta_2|^2(0)}}\right)$ is full elliptic integral of the first kind, $\Delta_1(0)$ and $\Delta_2(0)$ are values of the order parameters at $T=0$.
In the vicinity of $T_c$ condition (A6) transforms to:

$$\text{sgn}\left(-\frac{\lambda_{12}}{\lambda_{11}\lambda_{22}-\lambda_{12}\lambda_{21}}+\frac{\pi}{2T_c}\gamma_{12}\right)=\pm 1. \quad (A8)$$